\documentclass{emulateapj}
\usepackage{amsmath}
\usepackage{graphicx}
\usepackage{float}
\usepackage{amsmath}
\usepackage{epsfig,floatflt}
\usepackage{subfigure}

\shorttitle{Progenitor of SN 2004dj}
\shortauthors{Wang et al.}

\def\gsim{\;\lower4pt\hbox{${\buildrel\displaystyle >\over\sim}$}\;}
\def\lsim{\;\lower4pt\hbox{${\buildrel\displaystyle <\over\sim}$}\;}
\def\grls{\;\lower4pt\hbox{${\buildrel\displaystyle >\over <}$}\;}

\begin{document}


\title{The Progenitor of Supernova 2004dj in a Star Cluster}


\author{Xiaofeng Wang\altaffilmark{1,2},
Yanbin Yang\altaffilmark{1}, Tianmeng Zhang\altaffilmark{1},
Jun Ma\altaffilmark{1}, Xu Zhou\altaffilmark{1}, Weidong
Li\altaffilmark{3},\\
Yu-Qing Lou\altaffilmark{1,2,4}, Zongwei Li\altaffilmark{5} }
\altaffiltext{1}{National Astronomical Observatories of China
(NAOC), Chinese Academy of Sciences (CAS), Beijing 100012, China}
\altaffiltext{2}{Physics Department and Tsinghua Center for
Astrophysics (THCA), Tsinghua University, Beijing, 100084, China;
wang\_xf@tsinghua.edu.cn, louyq@tsinghua.edu.cn}
\altaffiltext{3}{Department of Astronomy, UC Berkeley, CA
94720-3411, USA} \altaffiltext{4}{Department of Astronomy and
Astrophysics, the University of Chicago, 5640 South Ellis Avenue,
Chicago, IL 60637, USA} \altaffiltext{5}{Dept of Astronomy,
Beijing Normal U., Beijing, 100875, China}



\begin{abstract}
The progenitor of type II-plateau supernova (SN) 2004dj is
identified with a supergiant in a compact star cluster known as
"Sandage Star 96" (S96) in the nearby spiral galaxy NGC 2403,
which was fortuitously imaged as part of the
Beijing-Arizona-Taiwan-Connecticut (BATC) Multicolor Sky Survey
from Feb 1995 to Dec 2003 prior to SN 2004dj. The superior
photometry of BATC images for S96, taken with 14 intermediate-band
filters covering $3000-10000$\AA, unambiguously establishes the
star cluster nature of S96 with an age of $\sim 20$Myr, a
reddening of $\hbox{E}(B-V)\sim 0.35$ mag and a total mass of
$\sim 96,000$M$_{\odot}$. The compact star cluster nature of S96
is also consistent with the lack of light variations in the past
decade. The SN progenitor is estimated to have a main-sequence
mass of $\sim$12M$_{\odot}$. The comparison of our
intermediate-band data of S96 with the post-outburst photometry
obtained as the SN has significantly dimmed, may hopefully
conclusively establish the nature of the progenitor.
\end{abstract}



\keywords {galaxies: individual (NGC 2403) --- galaxies:
star clusters --- stars: evolution --- supergiants ---
supernovae: general --- supernovae: individual (SN 2004dj)}


\section{Introduction}
Identification of the progenitors of SNe is extremely valuable for
testing theories of stellar evolution and supernova explosions.
Type II SNe arise from core-collapses of evolved massive stars
(Paczynski 1971; Goldreich \& Weber 1980). The subclass of type
II-P SNe is thought to be associated with red supergiants of
higher initial masses that retained their hydrogen envelopes
before core collapse. This model accounts for the main observed
features (spectra and light curves) and the estimated physical
parameters of the expanding photosphere such as velocity,
temperature and density (Woosley \& Weaver 1986; Hamuy 2003).
However, there is little direct evidence for the red supergiant
hypothesis. To date, only three SNe have had their progenitors
identified: SN 1987A (Gilmozzi et al. 1987; Sonneborn et al.
1987), SN 1993J (Aldering et al. 1994; Maund et al. 2004), and SN
2003gd (Van Dyk et al. 2003; Smartt et al. 2004). The progenitor
of peculiar type II SN 1987A was a blue supergiant. The Type IIb
SN 1993J arose in a massive interacting binary. The progenitor of
type II-P SN 2004et in NGC 6946 was tentatively identified as a
yellow supergiant (Li et al. 2005). Thus the expected red
supergiant origin for the common type II-P SNe is only favored for
SN 2003gd. The fortuitous occurrence of type II-P SN 2004dj in a
nearby galaxy with prediscovery images has allowed us to examine
its progenitor.

SN 2004dj was discovered on July 31.76 UT 2004 by K. Itagaki
(Nakano 2004) in the nearby spiral galaxy NGC 2403 about $3.3$ Mpc
away (Karachentsev et al. 2004). Initially, the reported $V-$band
magnitude was $\sim 11.2$ mag, making SN 2004dj the optically
brightest SN in the past decade since SN 1993J in M81. The initial
spectrum of SN 2004dj resembles that of a normal type II-P
supernova, with prominent P-Cygni profiles in hydrogen Balmer
lines (Patat et al. 2004).


Through astrometric registration of archival images of NGC 2403
and recent images of SN 2004dj, Ma\'iz-Apell\'aniz et al. (2004;
MA04) established that SN 2004dj coincided with Sandage Star 96
(S96) in the list of luminous stars and clusters in NGC 2403
(Sandage 1984). Using the earlier photometry of S96 (Larsen 1999),
MA04 suggested that S96 was a young compact cluster with an age of
13.6 Myr and a total stellar mass of 24,000 $M_{\odot}$. They
inferred that the progenitor of SN 2004dj had a main-sequence mass
of $\sim 15M_\odot$.

In this Letter, we identify the progenitor of SN 2004dj
by combining recent and archival images from the
Beijing-Arizona-Taiwan-Connecticut (BATC) Sky Survey.
We find that the spectral energy distribution (SED) of
S96 resembles that of a star cluster rather than a single
star. Using the simple stellar population (SSP) model,
we re-examine the age and mass of S96 and set new limits
on the progenitor mass of SN 2004dj.

\section{BATC Observations of Sandage Star 96}

\subsection{Archival Images of the BATC Sky Survey}

The observations of NGC 2403 were obtained by the BATC 60/90cm
Schmidt telescope located at the XingLong station of the National
Astronomical Observatory of China (NAOC). This telescope has 15
intermediate-band filters covering the optical wavelength range of
$3000-10000\AA$, and is specifically designed to avoid
contaminations from the brightest and most variable night-sky
emission lines. Descriptions of the BATC photometric system can be
found in Fan et al. (1996).

Figure 1 compares the pre- and post-explosion images of SN 2004dj.
The left panel shows the field of SN 2004dj and neighboring bright
field stars, as imaged in the $i$ band (centered on 6656 \AA) of
the BATC system with the NAOC 60/90cm Schmidt telescope on 2004
Aug 11. In the right panel, we show the same field, as extracted
from the BATC $i$-band images taken on 1995 Dec 20. SN 2004dj
coincides with S96 to within 0.7$^{\prime\prime}$ based on
astrometry measurements in these two images.
\begin{figure}[htbp]
\figurenum{1}\hspace{-0.0cm}\plottwo{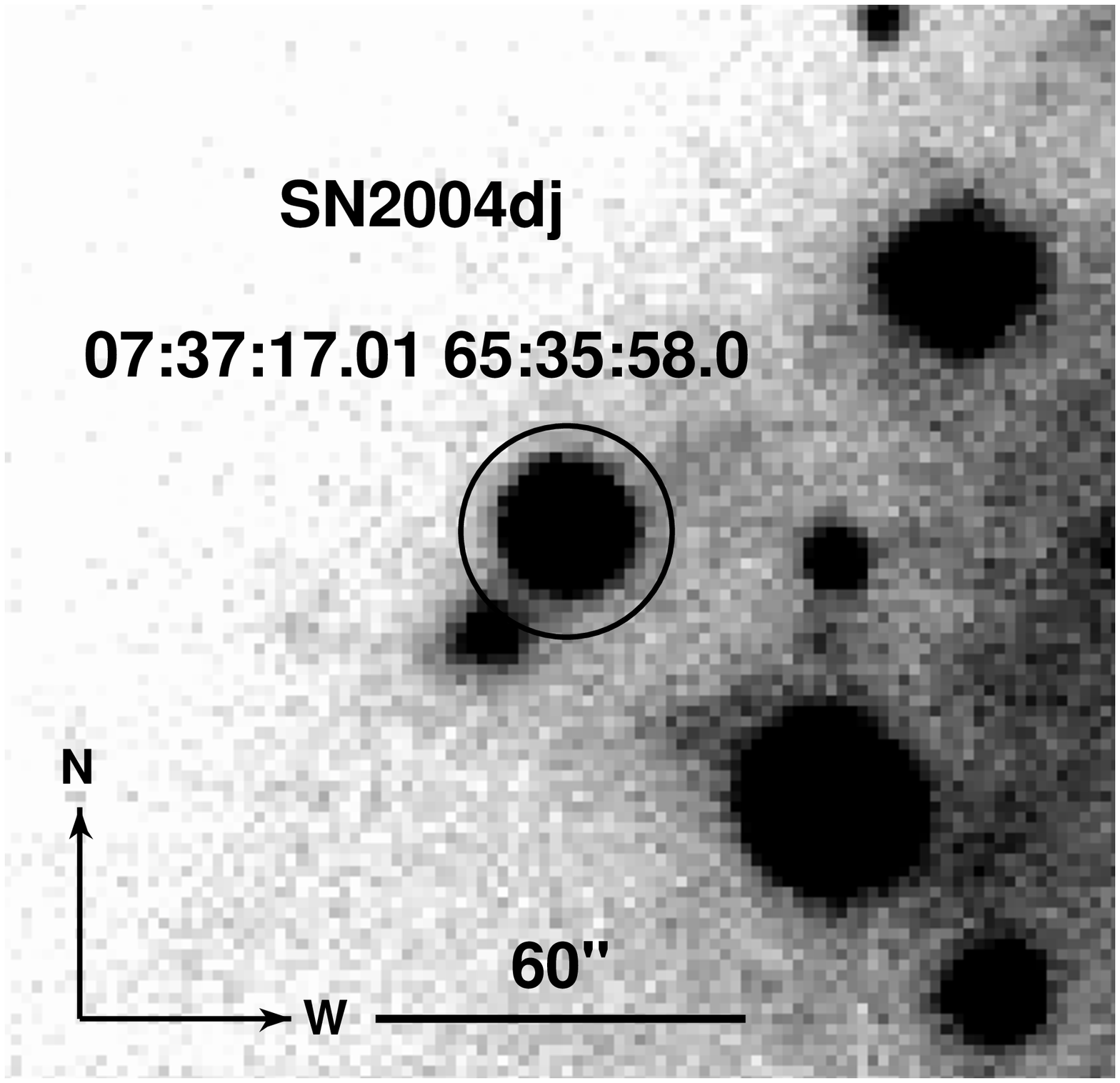}{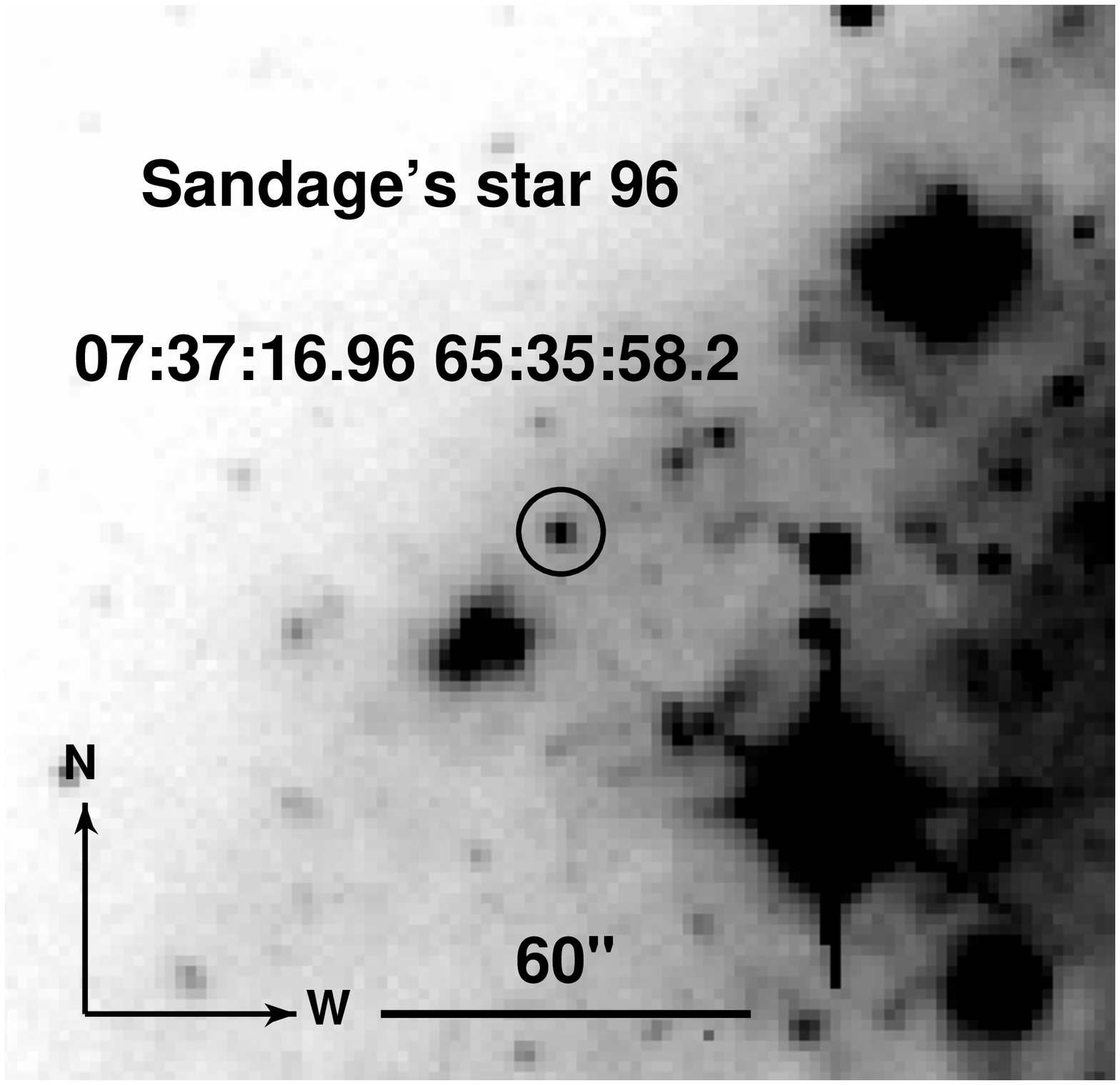}
\caption{(Left) An image of SN 2004dj in the BATC $i$ band of the
NAOC 60/90cm Schmidt telescope on 2004 Aug 11. The SN is circled.
(Right) The same field is observed before SN 2004dj on 1995 Dec
20. The SN is found to coincide with Sandage Star 96 or S96
(circled). The field of view of both images is $3^{\prime}\times
3^{\prime}$.} \label{fig:one}
\end{figure}
We extracted 213 images of NGC 2403, taken in 14 BATC filters
except for the $b$ filter, from the BATC survey archive during Feb
1995 to Dec 2003. Table 1 contains the log of observations.
Multiple images of the same filter were combined to improve the
image quality assuming a nonvariable S96. These serendipitous data
provide a unique opportunity to examine pre-SN S96.

\begin{table}
 \Large \caption{BATC Photometry of Sandage
Star 96} \vskip 0.1cm \hspace{0.0cm} \footnotesize
\begin{tabular}{cccccc}
\tableline \tableline
Filter&$\lambda$(\AA)&FWHM(\AA)&$N^{a}$&Value(mag)&Time Span\\
\tableline
   a &3360&360&8 & 18.73(0.24)&Dec 1998 \\
   c &4210&320&8 & 18.18(0.04)&Dec 1998 - Nov 2003 \\
   d &4540&340&38& 18.16(0.03)&Dec 1998 - Nov 2003\\
   e &4925&390&17& 18.12(0.07)&Mar 1995 - Jan 2003\\
   f &5270&340&9 & 18.12(0.05)&Feb 1995 - Dec 1995\\
   g &5795&310&13& 17.93(0.04)&Feb 1995 - Jan 1996\\
   h &6075&310&12& 17.85(0.06)&Feb 1995 - Jan 1996\\
   i &6656&480&11& 17.78(0.06)&Feb 1995 - Jan 1999\\
   j &7057&300&10& 17.61(0.09)&Feb 1995 - Dec 2003\\
   k &7546&330&13& 17.57(0.06)&Feb 1995 - Mar 1996\\
   m &8023&260&17& 17.57(0.08)&Feb 1995 - Dec 2003\\
   n &8480&180&17& 17.51(0.18)&Feb 1995 - Oct 2000\\
   o &9182&260&17& 17.45(0.11)&Feb 1995 - Dec 2003\\
   p &9739&270&23& 17.42(0.25)&Feb 1995 - Feb 2000\\
\tableline
\end{tabular}
\vskip 0.4cm \tablenotetext{a}{$N$ is the number of images taken
by the BATC telescope.} \vspace{-0.3cm}
\end{table}
\subsection{Intermediate-Band Photometry of S96}

S96 is located in a somewhat complex background involving a spiral
arm and possibly surrounding H II regions. To obtain proper
photometry, we need to consider the flux contribution from the
host galaxy background underneath S96. We invoked a method that
assumes the spiral plane around S96 being stable and satisfying
the Laplace equation and solves for the flux distribution of the
spiral arm at the position of S96 (Zhang et al. 2004). The pure
flux of S96 can then be estimated by subtracting the host galaxy
contribution from the total flux.

The final magnitudes of S96 are determined on the subtracted
images with the standard aperture photometry. The BATC photometric
system calibrates the magnitude zero level similar to the
spectrophotometric AB magnitude system. For the flux calibration,
the Oke-Gunn primary flux standard stars HD 19445, HD 84937, BD
+26$^{\circ}$2606, and BD +17$^{\circ}$4708 were observed during
photometric nights (Yan et al. 2000). The results of
well-calibrated photometry of S96 in 14 filters are summarized in
the fifth column of Table 1. These intermediate-band magnitudes
(Fig. 3) agree well with the broad-band $UBVI$ magnitudes of
Larsen (1999).

The numerous BATC images of NGC 2403 (Table 1) allow us to examine
the variability of S96 in the past decade, which will shed light
on the nature of the object. For example, significant light
variations are expected on a time scale of a few years, if S96
were a luminous blue supergiant (Humphrey \& Davidson 1994). We
converted all BATC magnitudes, measured at different epochs and
with higher qualities, into the $i$ band by utilizing the color of
S96 inferred from Table 1. As shown in Fig 2, the composite
$i$-band light curve of S96 in the past decade does not show
significant luminosity variations and most of the data are within
$\pm 0.13$ mag (1$\sigma$) of the average value. The fit for a
flat light curve gives $\chi^{2}$ = 49 for 43 data points. This
also justifies that the image combination employed in our data
reduction is reasonable.

\begin{figure}[htbp]
\figurenum{2}\vspace{-0.0cm}\plotone{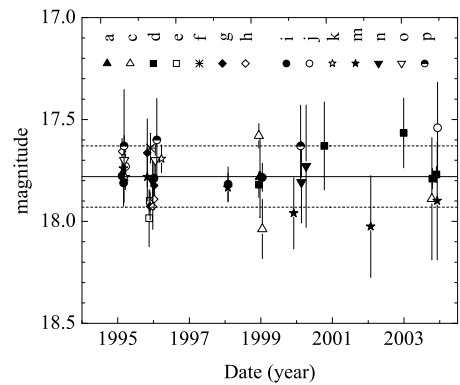}\caption{The composite
$i$-band light curve of S96. Solid circles show the $i$ band
magnitudes and other symbols are those converted from other 13
bands. All BATC band symbols are labeled on top. The solid line is
the mean value of the $i$-band magnitudes and the two horizontal
dashed lines mark $\pm 1\sigma$ error.}
\label{fig:two}\vspace{-0.2cm}
\end{figure}

\section{Stellar Population of Sandage Star 96}

\subsection{Stellar Populations and Synthetic Photometry}

To probe the nature of S96, we compare its SED with theoretical
stellar population synthesis models. We explore two different SED
families, one for single stars and the other for cluster
populations. For stellar models, we used the theoretical stellar
spectral flux library of Lejeune et al. (1997), covering a wide
range of stellar parameters. For cluster spectra, we used the SSP
models (e.g., Bruzual \& Charlot 2003; BC03),
which assumed a single generation of coeval stars with fixed
parameters such as age, metallicity and initial mass function.

We convolve the SEDs of single stars from Lejeune et al. (1997)
and stellar clusters from BC03 with the BATC filter transimision
curves to obtain synthesized optical and near-infrared photometry
for comparisons. The synthetic $i$-th BATC filter magnitude can be
computed by
\begin{equation}
m=-2.5\log\frac{\int_{\lambda}F_{\lambda}\varphi_{i}
(\lambda)d\lambda}{\varphi_{i}(\lambda)d\lambda}-48.60
\end{equation}
where $F_{\lambda}$ is the SED from the library and $\varphi_{i}$
is the transmission function of the $i$-th filter of the BATC
photometric system. Here, $F_{\lambda}$ varies with temperature
and metallicity for stellar models, and with age and metallicity
for star clusters. We explore the nature of S96 by fitting the
observed SED with distinct theoretical models.

\subsection{Reddening and Metallicity of S96}

To obtain the intrinsic SED of S96, the photometry must be
de-reddened. The Na ID interstellar absorption lines can provide
clues to the reddening. However, the two reports about the
measurements of these feature lines of SN 2004dj (Patat et al.
2004; Gunther et al. 2004) show larger difference which prevents
confident determination of the reddenings towards S96. As a
result, we treat E$(B-V)$ as a fitting parameter determined
simultaneously along with the effective temperature for stellar
models or the age for cluster models. The values of extinction
coefficient R$_{\lambda}$ for the BATC filters are obtained by
interpolating the interstellar extinction curve of Cardelli et al.
(1989).


The SEDs for clusters or single stars are significantly affected
by the adopted metallicity. Garnett et al. (1997) measured the
metallicity and its radial gradient in NGC 2403, using an expanded
sample of H II regions. S96 is 160$^{\prime\prime}$ east and
10$^{\prime\prime}$ north of the galactic nucleus of NGC 2403, or
at 4.0 kpc from the galactic center when removing the projection
effect. At this radial distance, the relative oxygen abundance
log(O/H)+12 is $8.45\pm 0.10$ dex from Garnett et al., which is
$\sim$ 40\% of the solar value. We then adopt a subsolar
metallicity of $z=0.008$ for S96 in the analysis. \footnote{We
tested the validity of a presumed subsolar metallicity by treating
it as a free parameter. The best-fit metallicity of $0.012\pm
0.003$ is obtained by using the BATC photometry and 2MASS
JHK$_{s}$ data (with a reduced $\chi^2\sim 0.72$), which also
prefers a subsolar value. However, uncertainties in the fitting
for age and reddening increase significantly due to an additional
parameter.}.

\subsection{Fitting Results}
We use the $\chi^2$ test to examine which members of the
SED families are most compatible with the observed one.
We show in Fig. 3 the observed integrated photometry of
S96 and the best-fit SED model.

Our fitting results do not favor S96 as a single luminous star.
Fitting the BATC photometry for the stellar model yields the
solution of a highly reddened [E$(B - V)$ = 0.78$\pm$0.05 mag]
B2IV star, but its best fit $\chi^{2}$ is 24 for 14 BATC points
(with the DOF to be 12), indicating a rather poor quality of the
fit. There are another three reasons why this fit does not work
for S96: (1) the corrected intrinsic $B-V$ color is too blue to be
consistent with a B-type star; (2) such an extremely blue and
bright star is most likely to be a luminous blue variable, but it
did not show significant light variations (see Fig. 2); (3) the
high reddening would imply an unrealistically high luminosity,
$M_V \lesssim -18.9$ mag at the maximum light, for SN 2004dj as a
type II-P event (e.g., Hamuy 2003).

\begin{figure}[htbp]
\figurenum{3}\vspace{-3.0cm}\hspace{-0.5cm} \plotone{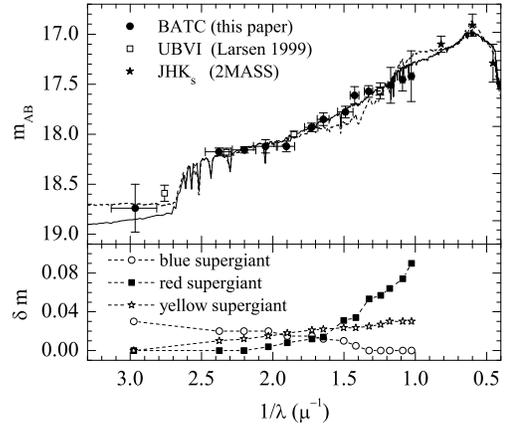}
\vspace{-1.8cm} \caption{Upper panel: the best-fit SED (20 Myr
solution) for S96, overlaid with that obtained by MA04 (dashed).
The photometric points are shown with error bars (vertical ones
for uncertainties and horizonal ones for the FWHM of BATC filter).
Lower panel: predicted flux drops of S96 when the SN has faded
away.} \label{fig:three}
\end{figure}

In comparison, the star cluster models provide better fits to the
observed SED of S96. The best reduced $\chi^2$ of 0.76 is achieved
for a cluster model with an age of 19.1$\pm$3.3Myr and a reddening
of E$(B-V)$=0.34$\pm$0.05 mag. When the 2MASS infrared $JHK_{s}$
data (Skrutiskie et al. 1997) were included in the fit, the
best-fit parameters for S96 are a cluster age of 20.0$\pm$3.4Myr
and a reddening of E$(B-V)$=0.35$\pm$0.05 mag with a similarly
small reduced $\chi^{2}$ of 0.77. The reddenings yielded for these
solutions are actually compatible with the one inferred from the
reported Na ID feature lines (e.g., Patat et al. 2004), if the
calibration by Munari \& Zwitter (1997) is adopted. The
uncertainties (1$\sigma$) in age and reddening are derived from
the likelihood map that restricted by the $\chi^{2}$ distribution
in the age-extinction plane. We reproduce in Fig. 3 the SED based
on the parameters derived by MA04 (the dashed curve), which does
not fit well the BATC data (especially in the $i$, $j$ and $k$
bands) since the best fit gives $\chi^{2}$ = 19 for 14 points.

Adopting these age and extinction estimates, we determine the
total mass of the stellar population from the distance to NGC
2403 and the measured $V$ magnitudes (transformed from observed
magnitudes in the BATC bands). We inferred the total mass of
$\sim 96,000$ M$_{\odot}$ by comparing the measured $V$ band
luminosity with the theoretical mass-to-light ratios (BC03).
The turn-off mass is $\sim$12$M_{\odot}$ for a cluster age of
20 Myr. The parameters estimated for S96 are listed in Table 2.

\begin{table}
\Large \caption{Fitting Parameters for Sandage star
96}\footnotesize \vskip 0.2cm \hspace{0.8cm}
\begin{tabular}{ccc}
\tableline \tableline
&BATC & BATC+$JHK_{s}$\\
\tableline
Age (Myr)&$19.1\pm3.3$&$20.0\pm3.4$\\
E$(B-V)$ (mag)&$0.34\pm0.05$&$0.35\pm0.05$\\
Mass ($10^{3}M_{\odot})$&$93.5\pm14.8$&$96.2\pm15.3$\\
Reduced $\chi^{2}$&0.76&0.77\\
Turnoff Mass ($M_{\odot}$)&12.2&11.7\\
\tableline
\end{tabular}
\vskip 0.2cm
\end{table}

\section{Discussion and Conclusions}
MA04 estimated a younger age of $\sim 13.6$Myr and a lower
reddening\footnote{There is also a solution for an older age of
$28.8$Myr and a larger reddening of E$(B-V)$ = 0.28 mag with a
reduced $\chi^{2}$ of $\sim$ 1.74 in their analysis but is
considered less likely.} of E$(B-V)$ = 0.17 mag for S96, using the
broad-band $UBVI$ photometry of Larsen (1999). Our results differ
from those of MA04 at a confidence level of $\sim 2\sigma$, which
is primarily caused by the differences in metallicity and spectral
coverage. Given a lower metallicity of $z=0.008$ for S96, however,
the combination of $UBVI$ and $JHK_{s}$ photometry yields two
solutions in the age-reddening plane. The old solution of age
$\sim 19$ Myr with a reduced $\chi^{2}$ of 0.89 is similar to
ours, while the young one of age $\sim$10 Myr is equally
compelling and cannot be rejected due to a smaller $\chi^{2}$ of
1.02. The occurrence of multiple solutions may be related to the
lower spectral resolution of the broadband photometry. In
comparison, the BATC data constrain the SED better. For instance,
our $i$-band photometry could indicate that S96 was not an
H$_{\alpha}$-bright source and hence the young solution is
unlikely, but the existing $UBVI$ data alone fail to do that. The
effect of different SED models on the results may be small as
suggested by de Grijs et al. (2005). This is manifest in our data
analysis of S96 using both BC03 and Starburst99 models (Letherer
et al. 1999). At a fixed metallicity, the age difference derived
from these two models is within $2-3$ Myr and the extinction
difference is within 0.06 mag.

While the observed SED of S96 resembles that of a star cluster,
several photometric points shown in Fig. 3 do not fit well, e.g.
the $f$ and $j$ bands. This small deviation of $\sim 0.1$ mag from
the best-fit SED may be caused by the model uncertainties, the
photometric errors, or even flux modulations of the immediate
progenitor star. In the last case, we set limits on the possible
variability. Given the luminosity of S96 and that predicted from
the supergiant with a main-sequence mass of 12M$_{\odot}$ (which
is $-6.9\lsim M_{bol}\lsim -7.8$ mag; e.g., Smartt et al. 2003),
we estimate that the flux contribution of the progenitor to be 3 -
8\% of the whole cluster, depending on the supergiant types. By
relegating all scatters ($\sim 0.13$ mag) shown in Fig. 2 to the
progenitor itself, we place a rather crude upper limit of $\sim
1.5$ mag on its light variation.

The progenitor mass of SN 2004dj inferred in S96
($\sim$12$M_{\odot}$) is within the mass range found for other
type II-P SNe (Leonard et al. 2002; Van Dyk et al. 2003; Smartt et
al. 2003, 2004). While it remains uncertain whether the progenitor
of SN 2004dj was a blue, yellow, or red supergiant in the compact
cluster S96, a comparison of flux changes for pre- and
post-outburst phases of S96 in the blue and red bands may offer
key clues to distinguish the various scenarios. Given a blue or
yellow supergiant, the post-SN flux of S96 would change by $\sim
0.02-0.03$ mag (Fig 3). If the progenitor was a red supergiant,
the flux drop of S96 at NIR bands from $k$ to $p$ would reach
$\sim 0.07$ mag. This change might be detectable by scrutinizing
S96, preferably with the {\it Hubble Space Telescope}, when the
supernova becomes significantly dimmed several years later (Zhang
et al. 2004).

We conclude that SN 2004dj occurred in the young compact star
cluster S96 of age $\sim 20$Myr. The lack of light variations in
S96 supports this cluster identification. The progenitor of SN
2004dj is inferred to be a supergiant with a main-sequence mass of
$\sim 12$M$_{\odot}$. Post-SN observations of S96 and detailed
photometric and spectroscopic studies of the SN evolution may
further constrain the progenitor of SN 2004dj.

\acknowledgments We thank the two referees for their constructive
suggestions. This work has been supported by the National Science
Foundation of China (grants 10303002, 10473012, and NKBRSF TG
199075402). YQL has been supported in part by the ASCI Center for
Astrophysical Thermonuclear Flashes at the U. of Chicago under DoE
contract B341495, by the Special Funds for MSBSRPC,
by the Collaborative Research Fund from the
NSFC for YOOCS
(NSFC 10028306) at the NAOC,
by NSFC grant 10373009 at the Tsinghua U, and by the Yangtze
Endowment from the MoE at the Tsinghua U.









\end{document}